\documentclass[twoside,journey]{IEEEtran}
\usepackage{makecell}
\usepackage{hyperref}
\usepackage{array}
\usepackage{graphicx,amssymb,amsmath}
\usepackage{multicol}
\usepackage[noadjust]{cite}
\usepackage{setspace}
\usepackage{subfigure}
\usepackage{graphicx}
\usepackage{float}
\usepackage {url}
\usepackage{stfloats}
\usepackage{amsthm,pifont}
\usepackage{flushend}
\usepackage{cases,subeqnarray}
\usepackage{bm,multirow,bigstrut}
\usepackage{amsmath, amsthm, amssymb}
\usepackage{textcomp}
\usepackage{latexsym,bm}
\usepackage{booktabs}
\usepackage{xcolor}
\usepackage{mathtools}
\usepackage{dsfont}
\usepackage{extarrows}
\usepackage{epsfig}
\usepackage{epsfig}
\usepackage{epstopdf}
\usepackage[noend]{algpseudocode}
\usepackage{algorithmicx,algorithm}
\usepackage{svg}
\theoremstyle{plain}

\theoremstyle{plain}

\usepackage{amsmath}

\IEEEoverridecommandlockouts

\begin{document}


\title{Exploring Collaborative Distributed Diffusion-Based AI-Generated Content (AIGC) in Wireless Networks}
\author{Hongyang~Du, Ruichen Zhang, Dusit~Niyato,~\IEEEmembership{Fellow,~IEEE}, Jiawen~Kang, Zehui Xiong, Dong~In~Kim,~\emph{Fellow,~IEEE},
Xuemin~(Sherman)~Shen,~\IEEEmembership{Fellow,~IEEE}, and H.~Vincent~Poor, \emph{Life Fellow,~IEEE}
\thanks{For more technical details about this paper, please refer to {\textit{``User-Centric Interactive AI for Distributed Diffusion Model-based AI-Generated Content''}} in {\url{https://arxiv.org/abs/2311.11094}}.}
\thanks{H.~Du is with the School of Computer Science and Engineering, the Energy Research Institute @ NTU, Interdisciplinary Graduate Program, Nanyang Technological University, Singapore (e-mail: hongyang001@e.ntu.edu.sg). D. Niyato is with the School of Computer Science and Engineering, Nanyang Technological University, Singapore (e-mail: dniyato@ntu.edu.sg). R. C. Zhang is with the School of Computer and Information Technology, Beijing Jiaotong University, Beijing 100044, China (e-mail: ruichen.zhang@bjtu.edu.cn). J. Kang is with the School of Automation, Guangdong University of Technology, China. (e-mail: kavinkang@gdut.edu.cn). Z. Xiong is with the Pillar of Information Systems Technology and Design, Singapore University of Technology and Design, Singapore (e-mail: zehui\_xiong@sutd.edu.sg). D.~I.~Kim is with the Department of Electrical and Computer Engineering, Sungkyunkwan University, Suwon 16419, South Korea (email:dikim@skku.ac.kr). X. Shen is with the Department of Electrical and Computer Engineering, University of Waterloo, Canada (e-mail: sshen@uwaterloo.ca). H.~V.~Poor is with the Department of Electrical and Computer Engineering, Princeton University, Princeton, NJ 08544, USA (e-mail: poor@princeton.edu).}
}
\maketitle

\begin{abstract}
Driven by advances in generative artificial intelligence (AI) techniques and algorithms, the widespread adoption of AI-generated content (AIGC) has emerged, allowing for the generation of diverse and high-quality content. Especially, the diffusion model-based AIGC technique has been widely used to generate content in a variety of modalities. However, the real-world implementation of AIGC models, particularly on resource-constrained devices such as mobile phones, introduces significant challenges related to energy consumption and privacy concerns.
To further promote the realization of ubiquitous AIGC services, we propose a novel collaborative distributed diffusion-based AIGC framework. By capitalizing on collaboration among devices in wireless networks, the proposed framework facilitates the efficient execution of AIGC tasks, optimizing edge computation resource utilization. Furthermore, we examine the practical implementation of the denoising steps on mobile phones, the impact of the proposed approach on the wireless network-aided AIGC landscape, and the future opportunities associated with its real-world integration. The contributions of this paper not only offer a promising solution to the existing limitations of AIGC services but also pave the way for future research in device collaboration, resource optimization, and the seamless delivery of AIGC services across various devices. Our code is available at \url{https://github.com/HongyangDu/DistributedDiffusion}.
\end{abstract}
\IEEEpeerreviewmaketitle

\section{Introduction}\label{intro}
\IEEEPARstart{T}{he} ubiquity of Internet-enabled devices has increased demand for high-quality, readily available content. AI-Generated Content (AIGC) has emerged as a preferred approach, delivering personalized and dynamic content by applying artificial intelligence (AI) models~\cite{du2023generative}. Ross Goodwin's innovative novel ``1 the Road'' exemplifies AIGC's adaptability, ingeniously employing AI algorithms alongside sensor-equipped mobile devices to convert sensory data into a literary composition. Despite its promising potential, computational and storage limitations have constrained AIGC's creative scope. Nevertheless, the advent of state-of-the-art fifth-generation (5G) technology and high-performance computing systems has rendered AIGC indispensable for generating creative and intricate content. Two prime examples of AIGC's impact include OpenAI's ChatGPT~\cite{van2023chatgpt} and Meta AI's Segment Anything Model (SAM)~\cite{kirillov2023segany}. ChatGPT, an AI chatbot, gained 100 million active users within two months of its launch, marking it as the fastest-growing consumer application in history~\cite{van2023chatgpt}. On the other hand, SAM is a cutting-edge AI model that can ``cut out" any object in any image with a single click. Trained on a dataset of 11 million images and 1.1 billion masks, SAM exhibits robust zero-shot performance on a diverse range of segmentation tasks~\cite{kirillov2023segany}.

As the cornerstones of AIGC, generative AI techniques have been instrumental in expanding content generation services. In particular, the diffusion model has emerged as a versatile and promising approach. The diffusion model operates through a probabilistic process in which the AI system iteratively reconstructs the original data from a series of noise-infused versions~\cite{ho2020denoising}. This innovative approach allows the diffusion model to learn and capture intricate patterns and structures inherent in a wide array of content types, thereby enabling the creation of coherent, contextually relevant, and aesthetically appealing outputs. The flexibility of the diffusion model has led to its widespread adoption in various AIGC applications:
	\begin{itemize}
		\item {\textit{Vision:}} Excelling in diverse image and video generation tasks, diffusion models have become integral to vision applications, such as image inpainting and text-to-image generation. A notable example is Stability AI's Stable Diffusion~\cite{rombach2022high}, a deep learning text-to-image model developed in 2022.
		\item {\textit{Audio:}} In audio generation, diffusion models demonstrate versatility across different content domains. For instance, diffusion models have been used to create piano rolls by leveraging a binomial prior distribution~\cite{atassi2023generating}.
		\item {\textit{Natural Language:}} The applications of diffusion models to natural languages has attracted significant interest~\cite{gong2023sequence}. Due to the iterative reconstruction process in diffusion-based models, the diffusion model offer great flexibility and improved trade-offs between content quality and efficiency~\cite{gong2023sequence}.
		\item {\textit{Time Series:}} One such development is the application of deep diffusion models that generate synthetic electronic health records~\cite{he2023meddiff}. The synthetic samples can be used for methodological developments and training purposes without using real patients' private health information.
		\item {\textit{Decision-making:}} Diffusion models have been used in generating optimal decisions, showing potential in sequential decision-making and various problem-solving domains~\cite{ajay2022conditional}. They have also been applied to wireless networks as AI-generated incentive mechanisms~\cite{du2023ai} and integrated with deep reinforcement learning algorithms~\cite{du2023generative}.
\end{itemize}

Despite the remarkable advances in AIGC, real-world implementation on various devices poses numerous challenges that warrant further exploration. These challenges include the intricate process of training and deploying AIGC models and the computationally intensive nature of the inference stage~\cite{ho2020denoising}. One approach to address these concerns is deploying AIGC models on server-side infrastructure, effectively offloading the computational demands. However, this solution may not be universally appealing, as users may prefer performing AIGC tasks on their local or nearby devices, due to security and privacy considerations, and in potential applications of sensitive domains like healthcare.
Meanwhile, the limited computational resources of these devices introduce significant challenges, potentially impacting the generation and inference time of AIGC models~\cite{du2023generative}. This problem is particularly evident in diffusion model-based AIGC, where each denoising step necessitates substantial energy consumption~\cite{linsley2020stable,ho2020denoising}.
To date, most AIGC research has focused on developing models in isolated server environments, neglecting the potential benefits of device collaboration. In real-world scenarios, ensuring efficient and seamless access to AIGC services is of considerable importance, as it directly impacts user satisfaction. Therefore, it is important to investigate innovative methodologies that address the challenges stemming from the collaborative execution of AIGC tasks across a multitude of devices and the diverse access requirements of users.

Recognizing the challenges faced by AIGC service execution on resource-constrained devices, we propose a collaborative distributed diffusion-based AIGC framework to save computing energy and enhance user experience. Within this distributed computing framework, we implement an effective strategy that enables devices to collaborate in performing shared denoising steps. The shared denoising steps can be executed on one device, i.e., an edge server or end device. Upon completing the shared steps, the intermediate results are wirelessly transmitted to other devices, which subsequently conduct the remaining task-specific denoising steps. This distributed computing method can also be viewed as an offloading technique, addressing privacy concerns by empowering users to maintain control over their content while saving computational resources across the network.
Our contributions can be summarized as follows:
\begin{itemize}
\item We present an in-depth analysis of diffusion model-based AIGC, examining its potential deployment in wireless networks. We also explore the underlying principles that facilitate the partitioning of the diffusion process.
\item We propose a collaborative distributed diffusion model-based AIGC framework. By integrating central and edge inference, the collaborative distributed computing approach effectively addresses the computational resource limitations inherent in diffusion model-based AIGC models, providing an efficient, scalable, and personalized experience for users.
\item We demonstrate the successful implementation of the Stable Diffusion v1-4 Model~\cite{rombach2022high} on a mobile phone, operating without an Internet connection. We present a comprehensive discussion of the proposed framework, providing insights into its potential impact on the wireless network-empowered AIGC. 
\end{itemize}

\section{Collaborative Distributed Diffusion-Based AI-Generated Content (AIGC) Framework}
In this section, we introduce the basic principles of diffusion models and the distributed computing of the diffusion process. Then, we present the collaborative distributed diffusion-based AIGC framework.
\subsection{Overview of Diffusion Models}
AI models have demonstrated remarkable advances in generating visually impressive images based directly on text descriptions. Diffusion models are the primary methodology used in text-generated images and serves as the core technology for this task. Several well-known and popular text-generated image models, including Stable Diffusion, Disco-Diffusion, Mid-Journey, and DALL-E2, are based on the diffusion model~\cite{ho2020denoising}. Among these, Stable Diffusion stands out as a milestone in AI image generation, providing high-performance results with higher-quality images, faster computation, lower resource consumption, and a smaller memory footprint~\cite{linsley2020stable}. We then present the fundamental principles of diffusion models and explore the potential for offloading the diffusion process.

\subsubsection{Principles of Diffusion Models}
Diffusion models are advanced generative models designed to create data that closely resembles the input training data~\cite{linsley2020stable}. The key principles are:
\begin{itemize}
	\item Systematic degradation of training data: The models introduce Gaussian noise step-by-step to degrade the original data. This step is called {\textit{Diffusion}}.
	\item Restoration of original data: Diffusion models learn to restore the data by reversing the noising process through incremental denoising and reconstruction. This step is called {\textit{Denoising}}.
	\item Modeling complex data distributions: The core idea involves iteratively transforming a simple Gaussian distribution into the target distribution.
	\item Neural networks as denoising functions: Diffusion models use neural networks to capture intricate relationships within the data, enabling high-fidelity sample generation and improved data synthesis.
	\item Demonstrated success across applications: These models have achieved remarkable results in image synthesis, text generation, and reinforcement learning~\cite{ho2020denoising}.
\end{itemize}

\subsubsection{Workflow of Diffusion Model-based AIGC} In this section, we delineate the workflow for a diffusion model-based AIGC model. As shown in Fig.~\ref{difftheo}, we use Stable Diffusion~\cite{linsley2020stable,rombach2022high} as an example:
\begin{figure*}[t!]
	\centerline{\includegraphics[width=0.83\textwidth]{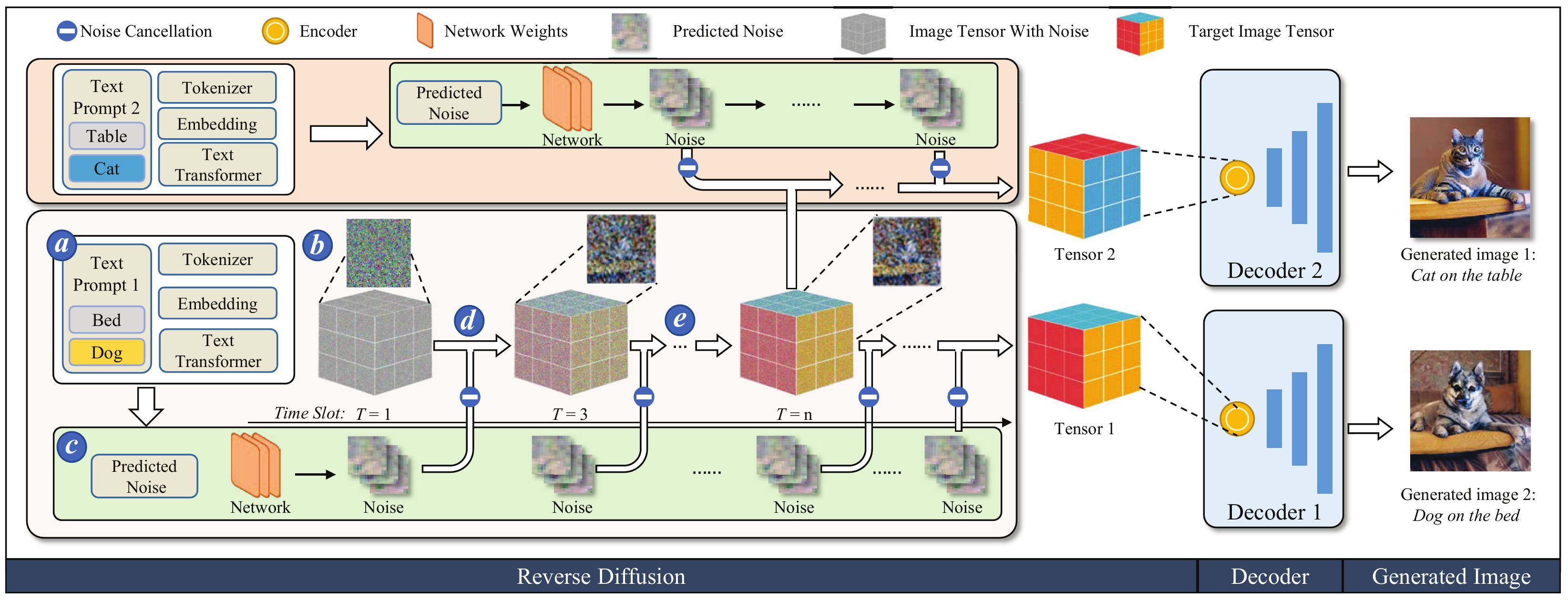}}
	\caption{The workflow of the diffusion model-based AIGC model, and the fundamental principles of implementing collaborative distributed AIGC. The text prompts for the two users are ``A bird on a table" and ``A cat on a table", respectively. Tokenizer transforms text into numerical tokens, enabling models to process complex phrases by breaking them into familiar parts. Embeddings represent words as vectors, capturing semantic relationships for tasks. Text transformer processes textual input data, conditioning the model for various natural language tasks and efficiently integrating text-based information.}
	\label{difftheo}
\end{figure*}

\textbf{a)} Text conditioning: This step involves the processing of textual prompts and their subsequent integration into a noise predictor. Within the Stable Diffusion framework, the textual prompt undergoes tokenization, is transformed into embeddings, and is subsequently processed by a text transformer before being utilized by the noise predictor~\cite{vaswani2017attention}. The text transformer serves the dual purpose of further refining the embeddings and offering a mechanism for incorporating various conditioning modalities. As a result, the output generated by the text transformer is utilized multiple times by the noise predictor, facilitated by a cross-attention mechanism.

\textbf{b)} Generation of a random latent tensor via Stable Diffusion: A random tensor is generated within the latent space. By configuring the seed for the random number generator, the stochastic nature of the tensor can be controlled, ensuring the reproduction of an identical random tensor when the same seed value is applied. Note that the tensor created at this stage is pure noise and does not correspond to any coherent image.

\textbf{c)} Application of the noise predictor: The noise predictor is a neural network that processes the input, which consists of the latent noisy image and a text prompt, and generates a prediction of the noise present within the latent space.

\textbf{d)} Computation of a new latent image tensor: The new latent image tensor is derived by subtracting the predicted noise tensor from the initial latent image tensor.

\textbf{e)} Iterative enhancement and final image generation: Utilizing the latent noisy image and noise prediction, Steps {\textbf{b)}} and {\textbf{c)}} are iteratively executed for a predefined number of sampling steps, resulting in the refinement of the image quality. Subsequently, the Variational Autoencoder (VAE) decoder~\cite{kingma2013auto} converts the improved latent image back into pixel space, producing the final image output.

\subsubsection{Wireless Network Architecture}
We explore different network architectures for the collaborative distributed diffusion-based AIGC system, each with its unique advantages:

\begin{itemize}
	\item \textbf{Edge-to-Multiple Devices:} This architecture employs an edge server as a communication hub for multiple user devices. The edge server performs shared denoising steps for groups of users with semantically similar task requirements and transmits the intermediate outputs to the respective user devices. The user devices then independently complete the remaining task-specific denoising steps. This architecture offers the following benefits:
	\begin{itemize}
		\item Reduced latency: By processing shared steps centrally, the architecture minimizes the time required for content generating and processing as the certain steps can be done on a more powerful computing  device.
		\item Optimized Resource Allocation: Centralized processing, with a complete view of the network load, allows for balanced task distribution across devices to enhance overall network efficiency.
		\item Effective load balancing across user devices: Distributing user-specific tasks prevents individual devices from becoming overloaded, ensuring a smooth user experience.
	\end{itemize}
	
	\item \textbf{Device-to-Device (Two Devices):} This architecture enables two devices to directly collaborate on distributed diffusion-based AIGC tasks. The devices jointly determine the shared denoising steps, perform these steps on a selected device, share intermediate outputs, and then individually complete the remaining denoising steps. This architecture provides the following advantages:
	\begin{itemize}
		\item Energy efficiency: Direct device-to-device communication bypasses the need for additional central processing, leading to energy savings.
		\item Enhanced privacy: Without the involvement of a central server, user devices independently execute their tasks, reducing the potential risk of AIGC content leakage.
	\end{itemize}
	
	\item \textbf{Forming a Cluster among Multiple Devices (with/without Edge):} In this architecture, user devices form collaborative clusters to jointly execute distributed AIGC tasks. Clustering can be accomplished either with the assistance of an edge server or through self-organization based on each device's task requirements and resources. These clusters collaboratively handle shared denoising steps, distribute intermediate outputs, and then individually complete the task-specific denoising steps. This architecture offers the following benefits:
	\begin{itemize}
		\item Adaptability: Clustering is inherently flexible, allowing the system to cater to a wide variety of AIGC task requirements by arranging devices based on their specific needs, such as one cluster for generating animal images and another for generating car images.
		\item Scalability: By dynamically adjusting cluster formations, this architecture can easily accommodate a growing number of devices and tasks.
		\item Resource optimization: By allowing devices to share computational resources and collectively perform AIGC tasks, clustering enhances overall system efficiency.
	\end{itemize}
\end{itemize}

\subsection{A Collaborative Distributed Diffusion-Based AIGC Approach}
\begin{figure*}[t!]
\centerline{\includegraphics[width=0.85\textwidth]{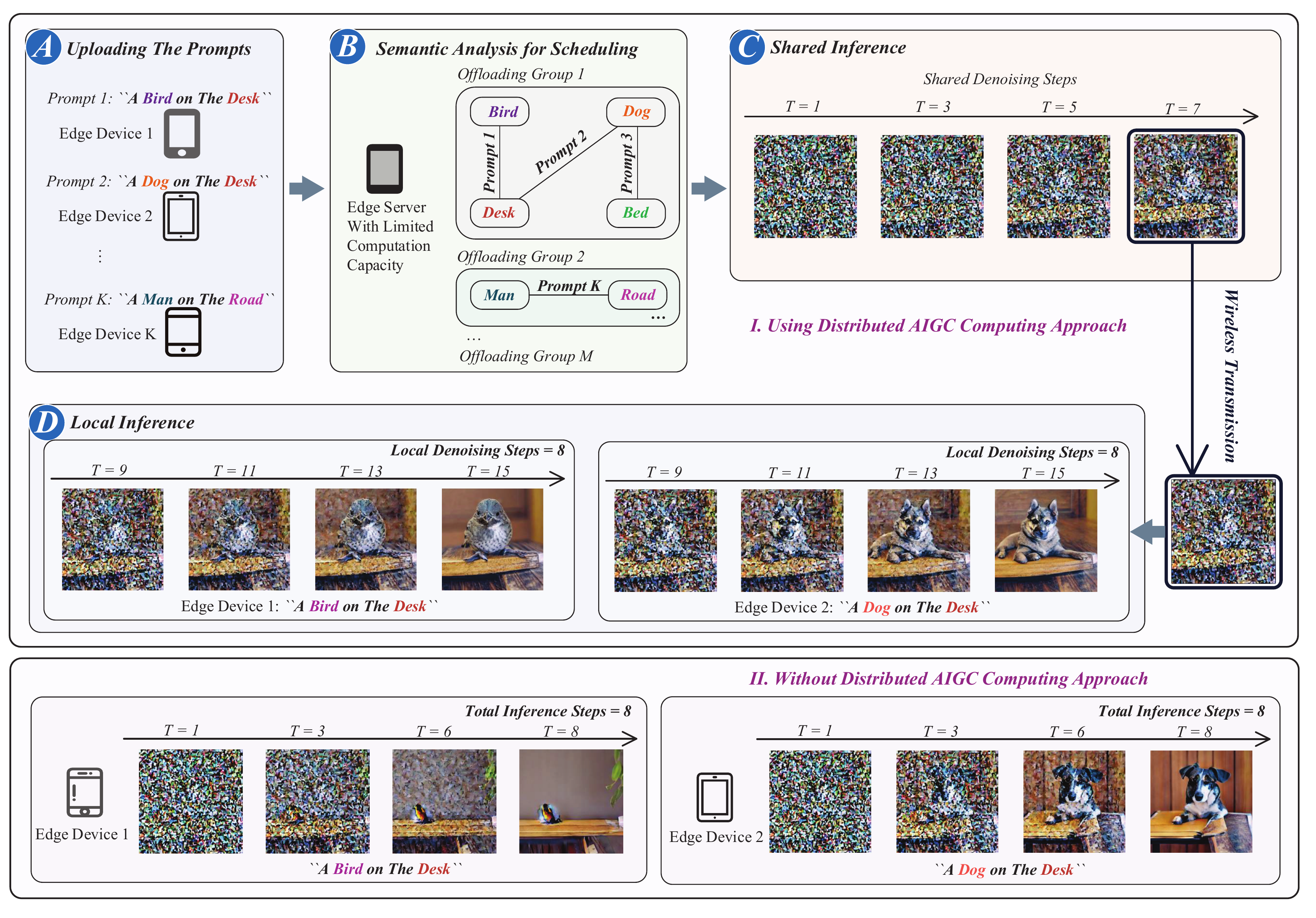}}
\caption{The proposed collaborative distributed diffusion-based AIGC approach in wireless networks. We consider the case where user end devices perform 8 local denoising steps and compare the results with those obtained when the collaborative distributed AIGC is not used.}
\label{Tutorial}
\end{figure*}
Given the network architecture, the collaborative distributed diffusion-based AIGC process can be executed through the following steps:

\textbf{Step 1. AIGC Model Training and Distribution:}
In the first step, AIGC models are trained using large datasets to ensure that they can generate high-quality content based on user inputs. The training process is conducted on high-performance computing devices, like GPU clusters, to handle the extensive computational requirements. Upon successful training, the models are distributed to edge servers, and they are also accessible for user devices to directly download.

\textbf{Step 2. Collect AIGC Task Requirements from Users:}
The system gathers AIGC task requirements from users, who submit requests containing textual prompts (e.g., ``A Bird on The Desk" as shown in Fig.~\ref{Tutorial} Part A) describing the desired content. Edge servers process and schedule these requests, ensuring efficient resource utilization and optimal system performance by understanding each task's specific requirements, including computational complexity and output quality.

\textbf{Step 3. Knowledge Graph-Aided Semantic Analysis and Offloading Scheduling:}
Upon collecting user requirements, the system conducts semantic analysis to discern similarities and differences between user prompts, facilitated by a knowledge graph~\cite{ji2021survey}. The graph (Fig.~\ref{Tutorial} Part B) offers a structured representation of semantic relationships, allowing the system to group users with similar task requirements and customize shared denoising steps for each group. Moreover, the graph can be updated incrementally, allowing for efficient handling of new tasks and facilitating frequent user reclustering.

\textbf{Step 4. Shared Inference:}
In the shared inference step, shared denoising steps (Fig.~\ref{Tutorial} Part C) are performed for each user group with similar task requirements on a central server. In this step, any text prompt in the grouped tasks can be used. The intermediate outputs after performing shared denoising steps are then transmitted to the respective edge devices, facilitating further processing. Note that our framework readily can integrate robust security measures, including data encryption techniques and physical layer security methods, in the data transmission process.

\textbf{Step 5. Local Inference:}
User devices receive intermediate outputs from the central server and proceed to complete user-specific denoising steps. By delegating these steps to user devices (Fig.~\ref{Tutorial} Part D), the system enables users to perform their tasks independently, conserving energy and maintaining privacy. As a result, users can efficiently generate the desired AI-generated content tailored to their requirements.

In summary, the proposed collaborative distributed diffusion-based AIGC framework aims to address the challenges and limitations of conventional AIGC systems. By performing shared denoising steps on a central server and offloading the remaining steps to edge devices, this approach balances computational load, reduces latency, and achieves high-quality content generation. Importantly, the versatility of this framework reaches beyond image-based AIGC, offering valuable potential solutions for a broad spectrum of diffusion model-based AI schemes in diverse domains as we discussed in Section~\ref{intro}.
Despite the advantages offered by this framework, it is crucial to validate its performance, practical implementation, and potential research challenges through numerical analysis and in-depth discussion in the following section.

\section{Numerical Results and Discussion}
In this section, we perform numerical results and discussion pertaining to collaborative distributed AIGC framework, organizing them into implementation and performance discussion.

\subsection{Implementation}
\begin{figure*}[t!]
	\centerline{\includegraphics[width=0.9\textwidth]{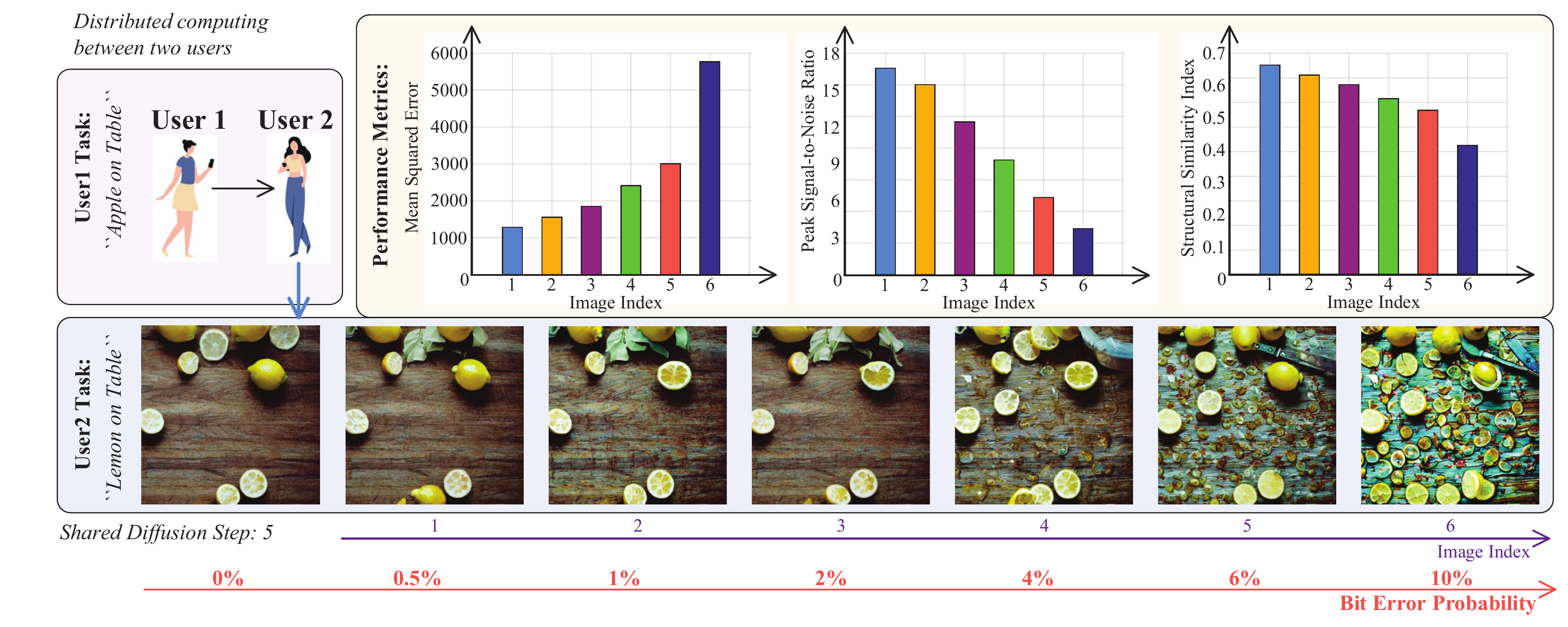}}
	\caption{The image quality metrics, i.e., MSE, PSNR, and SSIM, of the final generated images under various error rates, showing the impact of wireless transmission on the distributed AIGC computing approach. Specifically, text prompts for user 1 and user 2 are ``Apple on Table'' and ``Lemon on Table'', respectively. The used AIGC model is Stable Diffusion v1-4 Model~\cite{rombach2022high}. User 1's device performs 5 shared steps, and the intermediate results are transmitted to user 2. User 2 then performs 6 local steps, and the final generated image is displayed on user 2's device.}
	\label{zhongduan}
\end{figure*}
We implemented the Stable Diffusion v1-4 Model~\cite{rombach2022high} on a Redmi K40 smartphone, a device featuring 12GB of RAM, 256GB storage, and a Qualcomm Snapdragon 870. This model was operationalized without an Internet connection using the ncnn framework (\url{https://github.com/Tencent/ncnn}), optimized for mobile platforms. Our process involved transforming the CLIP (\url{https://github.com/openai/CLIP}) and diffusion model for offline deployment using the pnnx tool. We also employed EuLer-A (\url{https://huggingface.co/docs/diffusers/api/schedulers/euler_ancestral}) as the sampler during the prompt processing phase to ensure consistent output image resolution at $512*512$ pixels.

\begin{figure}[t!]
	\centerline{\includegraphics[width=0.43\textwidth]{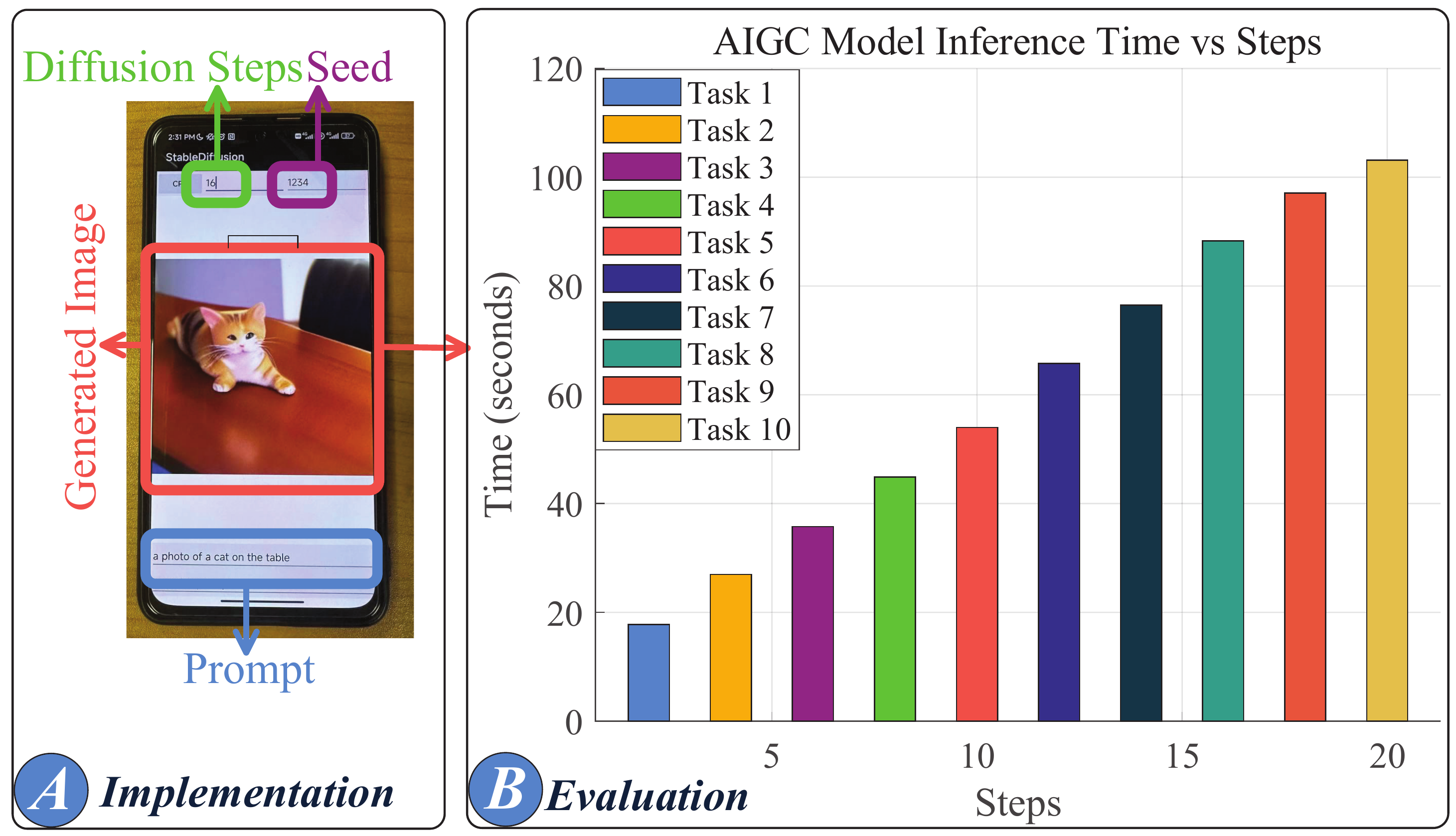}}
	\caption{The implementation of Stable Diffusion v1-4 Model~\cite{rombach2022high} in a mobile phone without the Internet connection, and the inference time test.}
	\label{bushu}
\end{figure}

{\textit{The feasibility of executing diffusion models on network end devices with limited computing power:}} We have implemented a diffusion model-based AIGC model, i.e., Stable Diffusion v1-4 Model~\cite{rombach2022high}, on a mobile phone. This successful deployment serves as empirical evidence of our proposed framework's practicality for local execution. Fig.~\ref{bushu} showcases the results obtained from this implementation. 
However, it is crucial to recognize that the inference speed and performance of diffusion models on mobile phones may be constrained by the device's computational capacity. This limitation can lead to extended processing times and potential restrictions on the complexity of the models and tasks that can be executed. To address these challenges, a range of optimization techniques can be employed, such as model pruning to reduce the model's size and complexity, quantization to decrease the numerical precision required for computations, and hardware acceleration to exploit specialized hardware components for improved performance.

{\textit{The effect of the wireless transmission on diffusion model-based:}} Wireless transmission can influence the performance of distributed diffusion model-based AIGC computing in edge networks. Several factors affect the successful and accurate transmission of diffusion results:
\begin{itemize}
	\item \textit{Transmit power:} Increasing transmit power enhances the signal-to-noise ratio (SNR), improving the accuracy and reliability of data transmission. Also, the transmit power can be adaptively allocated to the transmissions of diffusion tasks from different users given efficiency and fairness criteria.
	\item \textit{Fading:} Wireless channels often exhibit fading due to multipath propagation and shadowing. Fading can cause fluctuations in the received signal strength, leading to variations in the transmission quality, in which the transmission of diffusion tasks can be scheduled to avoid fading. For example, during deep fading, the edge server can perform more denoising steps and transmit the results to the mobile device once channel quality becomes better.
	\item \textit{Mobility:} The mobility of devices within an edge network may cause rapid changes in the channel conditions, requiring the communication system to adapt swiftly. High mobility may lead to increased handover frequency, resulting in temporary service disruptions and reduced system performance. Again, denoising steps can be adjusted according to the mobility patterns of different users.
\end{itemize}

We consider that when user 1 transmits the results after the shared denoising step to user 2, the wireless environment may introduce bit errors. Fig.~\ref{zhongduan} shows the generated image quality by user 2 after performing the local denoising steps using the received results from user 1, under different bit error probabilities. In addition to the visual presentation, we have calculated several image quality evaluation metrics, including mean squared error (MSE), peak signal-to-noise ratio (PSNR), and structural similarity index (SSIM)~\cite{mohammadi2015subjective}. We can observe that although the increase in bit error probability does cause damage to the final results, distributed AIGC computing has relatively high robustness. When the error rate reaches 2\%, user 2 can still produce high quality images. The reason is that the diffusion process carried out locally by user 2 can, to some extent, correct the image and improve the final generated quality due to the denoising step as shown in Fig.~\ref{difftheo}.

\subsection{Performance}
 We then discuss the concerns and considerations that arise when applying the proposed collaborative distributed AIGC framework to wireless networks.
{\textit{The impact of the proportion of joint denoising steps on system performance:}} In our proposed distributed AIGC computing framework, the proportion of shared denoising steps can significantly influence system performance. As the number of shared denoising steps increases, the system resources are more efficiently utilized. The reason is that multiple user tasks can share the same diffusion generation result, thereby reducing the workload for individual users and allowing for more effective use of available resources.
However, it is crucial to find a balance between the number of shared denoising steps and the quality of the generated images. If the number of denoising steps executed at the user's terminal for the own task is too small, generating high-quality images that meet the task requirements becomes challenging. This can lead to a trade-off between the proportion of shared denoising step and image quality.

\begin{figure}[t!]
\centerline{\includegraphics[width=0.43\textwidth]{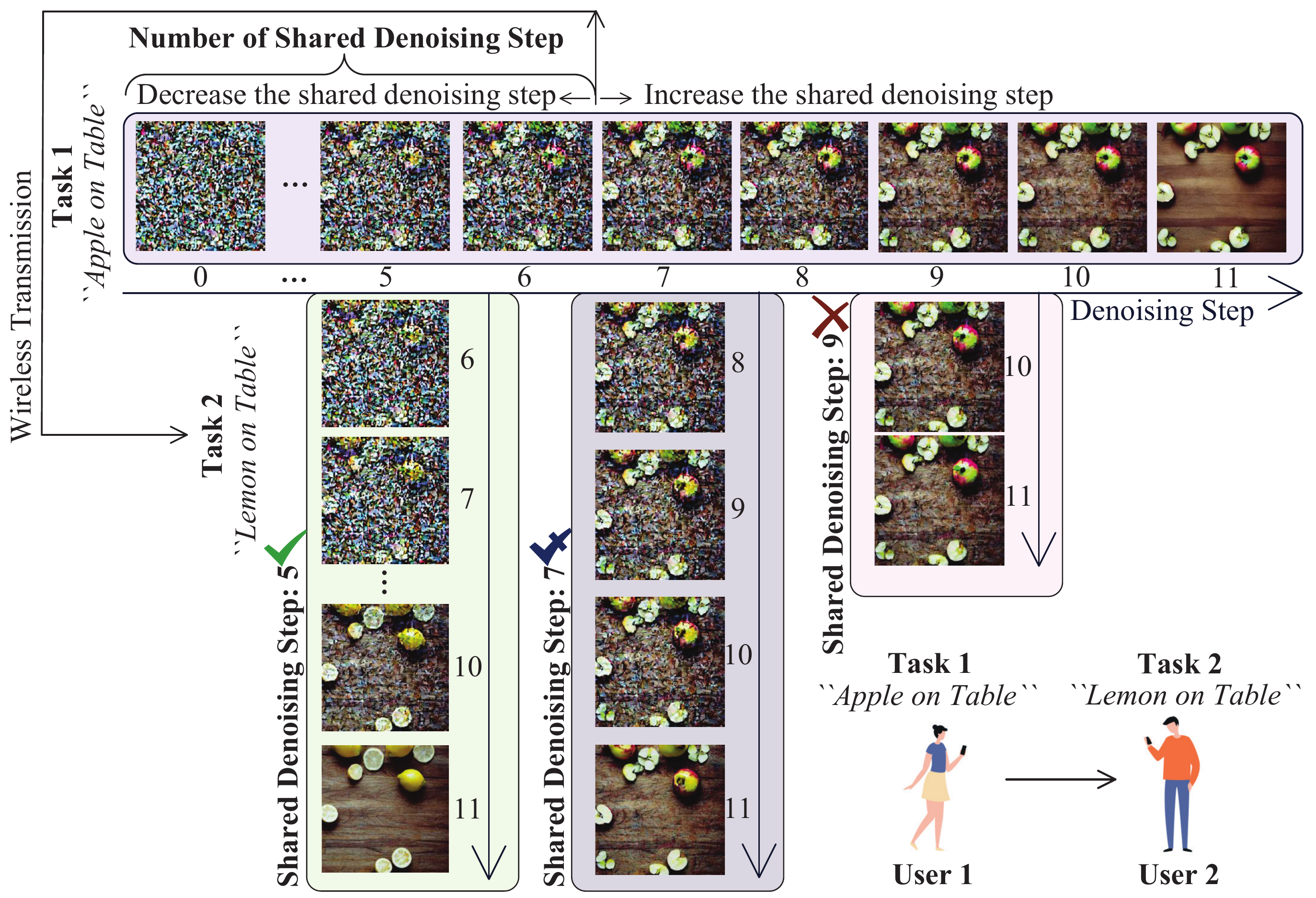}}
\caption{The impact of the proportion of shared denoising steps on system performance.}
\label{proportion}
\end{figure}
As shown in Fig.~\ref{proportion}, two users' prompts are ``{\textit{Apple on Table}}'' and ``{\textit{Lemon on Table}}'' respectively. User 1 executes the entire 11-step diffusion process. User 2 then receives the intermediate diffusion outcomes from user 1 to continue its own AIGC task. When the shared denoising steps are set to 5, user 2 can achieve a visually appealing result. However, as the number of shared denoising steps increases to 7, although the AIGC output still semantically meets user 2's prompt requirements, the generated image quality is negatively affected. When the shared denoising steps become excessive, such as 9 steps, user 2 only performs two local denoising steps, rendering it insufficiently to fulfill its own task request. Ideally, an optimal balance should be achieved to ensure efficient resource utilization without sacrificing image quality. In this context, our framework, when optimally configured with shared denoising steps, can outperform the centralized system while saving resources. This may involve conducting experiment and analysis to identify the most suitable balance for various use tasks.

\begin{figure}[t!]
\centerline{\includegraphics[width=0.43\textwidth]{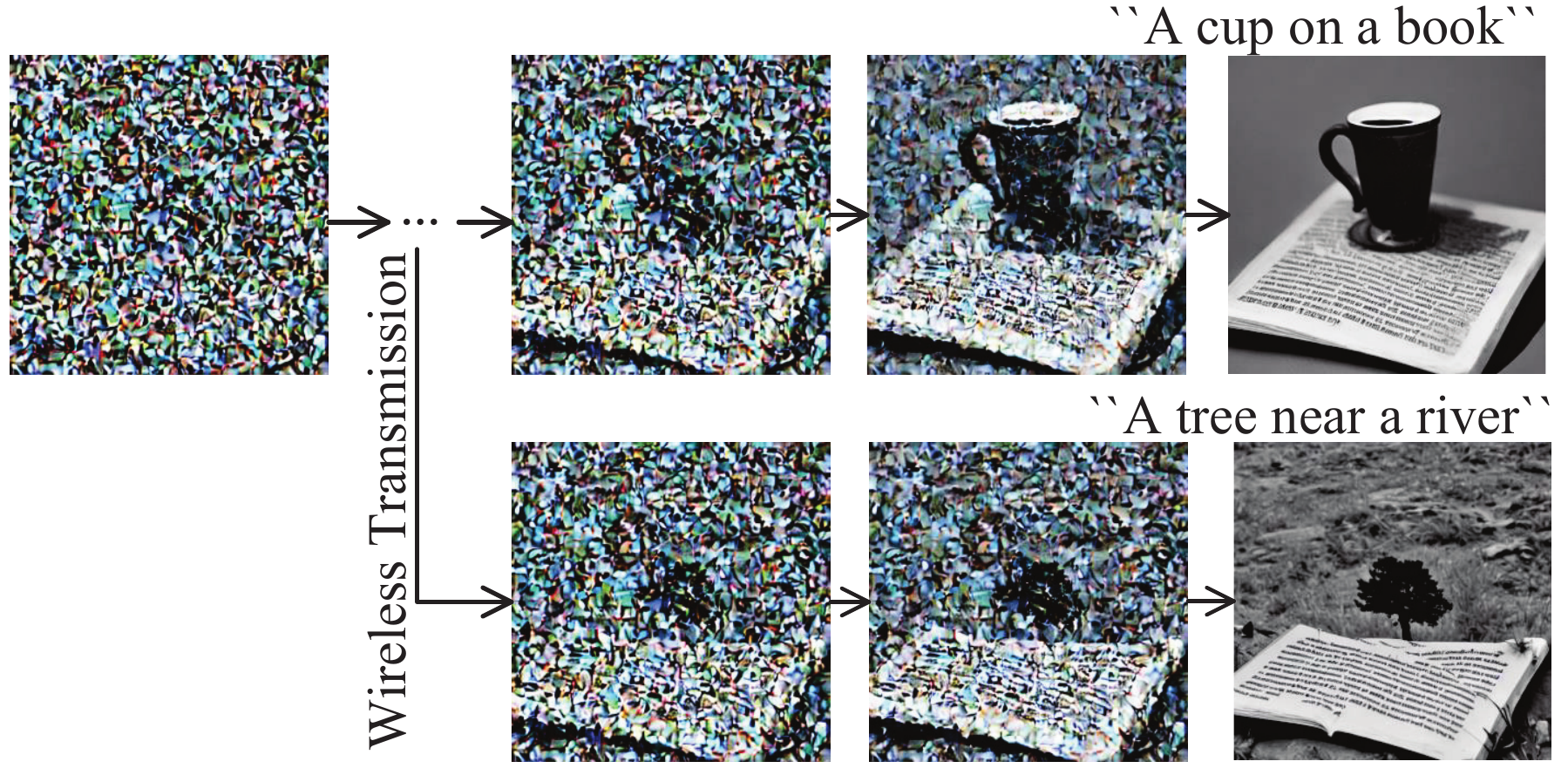}}
\caption{A failure example of the distributed AIGC computing. The numbers of total, shared, and local denoising steps are 11, 4, and 7, respectively.}
\label{differ}
\end{figure}
{\textit{The impact of variations in semantic content between users' prompts on the system performance:}} In the proposed distributed AIGC computing approach, a portion of shared denoising steps is executed initially, with the resulting intermediate outputs being transmitted to end devices to complete the remaining steps tailored to their specific tasks. When there is a significant variation in the semantic content of users' prompts, the efficacy of the shared denoising steps may be adversely affected, potentially leading to suboptimal final results. Fig.~\ref{differ} illustrates an example where distributed AIGC computing is ineffectively utilized due to the substantial difference in semantic content between users' prompts. 

To address this issue, it is essential to group users judiciously based on the semantic information of their tasks. By clustering users with similar semantic requirements, the shared denoising steps can be tailored to better serve the content demand of each group, ensuring a more effective transmission of intermediate results to end devices. Proper user clustering also contributes to the overall system performance by reducing the computational overhead and improving the efficiency of resource utilization. Moreover, a caching mechanism can be used as another solution. In this setup, the edge server stores or caches intermediate outputs from novel tasks, enabling faster and less resource-intensive processing for future tasks of similar semantic information.

\section{Future Directions}
\subsection{Incentive Mechanism Designs for Distributed AIGC Computing}
AIGC computing tasks in wireless networks require efficient completion to ensure optimal performance. Incentive mechanism design can enhance computing efficiency and reduce costs by motivating users and devices to contribute resources and participate in the computation process~\cite{du2023ai}. Although incentive mechanisms generally apply to various AI services, AIGC services present unique challenges due to their iterative and distributed nature, which necessitates close collaboration and synchronization among participating devices. As such, future research should focus on designing incentive mechanisms that consider the specific requirements and constraints of AIGC tasks, such as latency, synchronization, and resource availability, and on promoting resource sharing, cooperation, and the development of more efficient AIGC systems.

\subsection{Joint Diffusion and Channel Coding With Adaptive Modulation}
By jointly optimizing the diffusion model-based AIGC computing and channel coding, and tailoring the modulation and coding schemes to the prevailing channel conditions, the communication system can effectively optimize the balance between throughput and reliability. This approach involves designing both the diffusion model and channel coding to work in harmony, taking into account the specific characteristics of the AIGC and the channel conditions. To achieve this, the system can dynamically adapt to variations in channel quality, incorporating feedback mechanisms to ensure optimal AIGC performance.

\subsection{Secure Scheme Designs for Distributed AIGC Computing}
Privacy protection is a crucial aspect of AI services, including AIGC. Ensuring sensitive data remains secure during distributed AIGC computing is a promising research direction. In particular, incorporating blockchain-based technologies can help address these challenges by ensuring data decentralization and preventing malicious actors from disrupting the distributed AIGC computing process. For instance, using a blockchain-enabled consensus mechanism can maintain the integrity and confidentiality of intermediate AIGC results shared between devices during wireless transmission. In addition, investigating secure and effective methods for auditing and monitoring the transmission process, such as machine learning-based anomaly detection or cryptographic proofs of computation, can further ensure the correctness and security of the distributed AIGC computing process.


\section{Conclusion}
We have proposed a collaborative distributed AIGC computing approach to overcome the challenges related to diffusion model-based AIGC services on devices with limited computational resources. By capitalizing on the cooperative capabilities of devices, our distributed AIGC computing framework aims to enhance the overall efficiency and scalability of AIGC task execution.
We have delved into the distributed AIGC computing framework, elucidating its core principles, potential applicability in wireless networks, and the opportunities it creates for the consistent delivery of AIGC services across various devices. Moreover, we have engaged in numerical results analysis and discussion that investigate the practicality of our proposed approach, its influence on the AIGC ecosystem, and the hurdles associated with incorporating it into real-world scenarios.
As AIGC becomes an integral part of our digital lives, it is crucial to develop strategies capable of effectively catering to the growing demand for AIGC services. We hope that our work can serve as an inspiration for researchers and practitioners to further explore wireless network-empowered AIGC.

\bibliographystyle{IEEEtran}
\bibliography{Ref}

\end{document}